\documentclass[aps,pre,a4paper,10pt,twoside,twocolumn,showpacs,showkeys]{revtex4-1}

\usepackage{amssymb,bm,times,graphicx}

\usepackage{hyperref}
\hypersetup{unicode=true, pdftitle={The critical behavior of Stavskaya's probabilistic cellular automaton}, pdfauthor={J. Ricardo G. Mendonça}, pdfkeywords={Stavskaya model, probabilistic cellular automata, phase transition, Domany-Kinzel, directed percolation}, pdfsubject={Statistical Mechanics, Cellular Automata, Phase Transition}, pdfnewwindow=true, pdftoolbar=true, pdfmenubar=true, pdffitwindow=false, pdfstartview={FitH}, colorlinks=true, linkcolor=blue, citecolor=blue, urlcolor=blue, extension=pdf}

\def\leq{\leqslant}
\def\geq{\geqslant}

\begin{document}

\title{A Monte Carlo investigation of the critical behavior of Stavskaya's probabilistic cellular automaton}

\author{J. Ricardo G. Mendon\c{c}a}
\email[Email: ]{jricardo@usp.br}
\affiliation{Instituto de F\'{\i}sica, Universidade de S\~{a}o Paulo -- Caixa Postal 66318, 05314-970 S\~{a}o Paulo, SP, Brazil}

\begin{abstract}

Stavskaya's model is a one-dimensional probabilistic cellular automaton (PCA) introduced in the end of the 1960's as an example of a model displaying a nonequilibrium phase transition. Although its absorbing state phase transition is well understood nowadays, the model never received a full numerical treatment to investigate its critical behavior. In this brief report we characterize the critical behavior of Stavskaya's PCA by means of Monte Carlo simulations and finite-size scaling analysis. The critical exponents of the model are calculated and indicate that its phase transition belongs to the directed percolation universality class of critical behavior, as it would be expected on the basis of the directed percolation conjecture. We also explicitly establish the relationship of the model with the Domany-Kinzel PCA on its directed site percolation line, a connection that seems to have gone unnoticed in the literature so far.
\end{abstract}



\maketitle

% %% % %% % %% % %% % %% % %% % %% % %% % %% % %% % %% % %% % %% % %% % %% % %% % %% % %% % %% % 

\section{\label{intro}Introduction}

Stavskaya's model is a one-dimensional probabilistic cellular automaton (PCA) proposed in the end of the 1960's by the Russian school of Markov processes as an example of an interacting particle system presenting a nonequilibrium phase transition \cite{stavs68,shnirman,toom68}. The model is related with the directed site percolation (DP) process, of which it can be viewed as a one-sided version, as well as with the Domany-Kinzel PCA in one of its manifolds (cf.\ Sec.~\ref{model}) \cite{dkpca}. Roughly speaking, the phase transition in Stavskaya's model follows from its attractiveness (its tendency for forming clusters) and the existence of an absorbing state, and is well understood nowadays.

However, while many rigorous results exist for this model \cite{stavs68,shnirman,toom68,toom80,discrete,cmp91,dejong,gielis,depoorter,demaere}, it has never received a full numerical treatment to estimate its critical point and critical exponents. In this brief report we proceed to such an investigation of Stavskaya's model by Monte Carlo simulations and finite-size scaling techniques. Besides closing a gap in the characterization of the model, our results add another bit of evidence in favor of the DP conjecture, according to which phase transitions into an absorbing state in short-ranged single component systems in the absence of conserved quantities all belong to the same universality class of critical behavior \cite{janssen,grassb}. Remind that although the DP conjecture is grounded on solid theoretical arguments and has been verified in a host of model systems, it ({\it i\/}) could not be proved rigorously yet, and ({\it ii\/}) has only very thin experimental evidence \cite{expdp,takeuchi}, so that it continues to rely on model systems to sustain itself.

% %% % %% % %% % %% % %% % %% % %% % %% % %% % %% % %% % %% % %% % %% % %% % %% % %% % %% % %% % 

\section{\label{model}Stavskaya's model}

Stavskaya's model is a two-state PCA defined on a one-dimensional periodic lattice of $L$ cells specified by the configuration $\bm{\eta}(t) = (\eta_{1}(t),$ $\eta_{2}(t),$ $\ldots,$ $\eta_{L}(t)) \in \{0,1\}^{L}$ and evolving in discrete time $t \in \mathbb{N}$ according to the following very simple rule: with probability $\varepsilon \in [0,\,1]$, $\eta_{i}(t+1) = 1$, otherwise $\eta_{i}(t+1) = \eta_{i-1}(t) \cdot \eta_{i}(t)$.

Clearly, $\bm{1} = (1,1,\ldots,1)$ is an absorbing state of the model. It can be proven that there exists a critical $\varepsilon^{*}$ such that for $\varepsilon > \varepsilon^{*}$ the only invariant measure of Stavskaya's PCA is $\delta_{\bm{1}}$, the measure concentrated in $\bm{1}$, and that for $\varepsilon < \varepsilon^{*}$ the invariant measures are translation-invariant convex combinations of the form $\alpha\mu_{\varepsilon} + (1-\alpha)\delta_{\bm{1}}$, with $0 < \alpha < 1$ and $\mu_{\varepsilon}$ the measure that puts mass on configurations with density $0 < \mu_{\varepsilon}(1) < 1$ \cite{toom68,shnirman,discrete}. Early bounds on the critical point estimate $0.09 < \varepsilon^{*} < 0.323$ \cite{toom68,discrete}. The upper bound was eventually confirmed, but not improved, by different techniques \cite{dejong}, while the lower bound never received a reassessment; it should be remarked that lower bounds on critical values of interacting particle systems are notoriously difficult to obtain.

Stavskaya's PCA is related with the Domany-Kinzel (DK) PCA \cite{dkpca} by taking the complementary (negated) variables $\bar{\eta}_{i} = 1-\eta_{i}$. It can then be seen that Stavskaya's PCA corresponds to the DK PCA on the line $p_{1} = p_{2} = 1-\varepsilon$, i.e., over the directed site percolation (site DP) line of the DK PCA parameter space. Notice, however, that the dynamics in the DK PCA is defined for each of its two sublattices in the time direction (even and odd time steps), while the dynamics in Stavskaya's PCA is direct. On the site DP line, the DK PCA displays an inactive-active phase transition at the critical point $p_{1}^{*} = p_{2}^{*} = 0.705\,489(4)$ \cite{onody}, corresponding to $\varepsilon^{*} = 0.294\,511(4)$, within the rigorous bounds mentioned before (the numbers between parentheses indicate the uncertainty in the last digit or digits of the data). Curiously, the relationship between Stavskaya's PCA and the DK PCA seems to have gone unnoticed in previous investigations \cite{discrete,cmp91,dejong,gielis,depoorter,demaere}, although a coupling scheme with an ``independent oriented percolation'' process equivalent with site DP was used in \cite{dejong}. It is worth mentioning that Stavskaya's model, together with another PCA introduced by the same epoch, Vasil'ev's model \cite{discrete,vasil}---which corresponds to the $p_{2}=0$ line in the DK PCA or, equivalently, to a probabilistic version of CA rule 18 in Wolfram's classification scheme \cite{wolfram}---predates the DK PCA and related models by almost two decades, but did not receive much attention, not even when CA and PCA reentered the mainstream scientific agenda in the 1980's.

% %% % %% % %% % %% % %% % %% % %% % %% % %% % %% % %% % %% % %% % %% % %% % %% % %% % %% % %% % 

\section{\label{behavior}The critical behavior}

\begin{figure}
\centering
\includegraphics[viewport=95 226 426 719, scale=0.42, angle=-90]{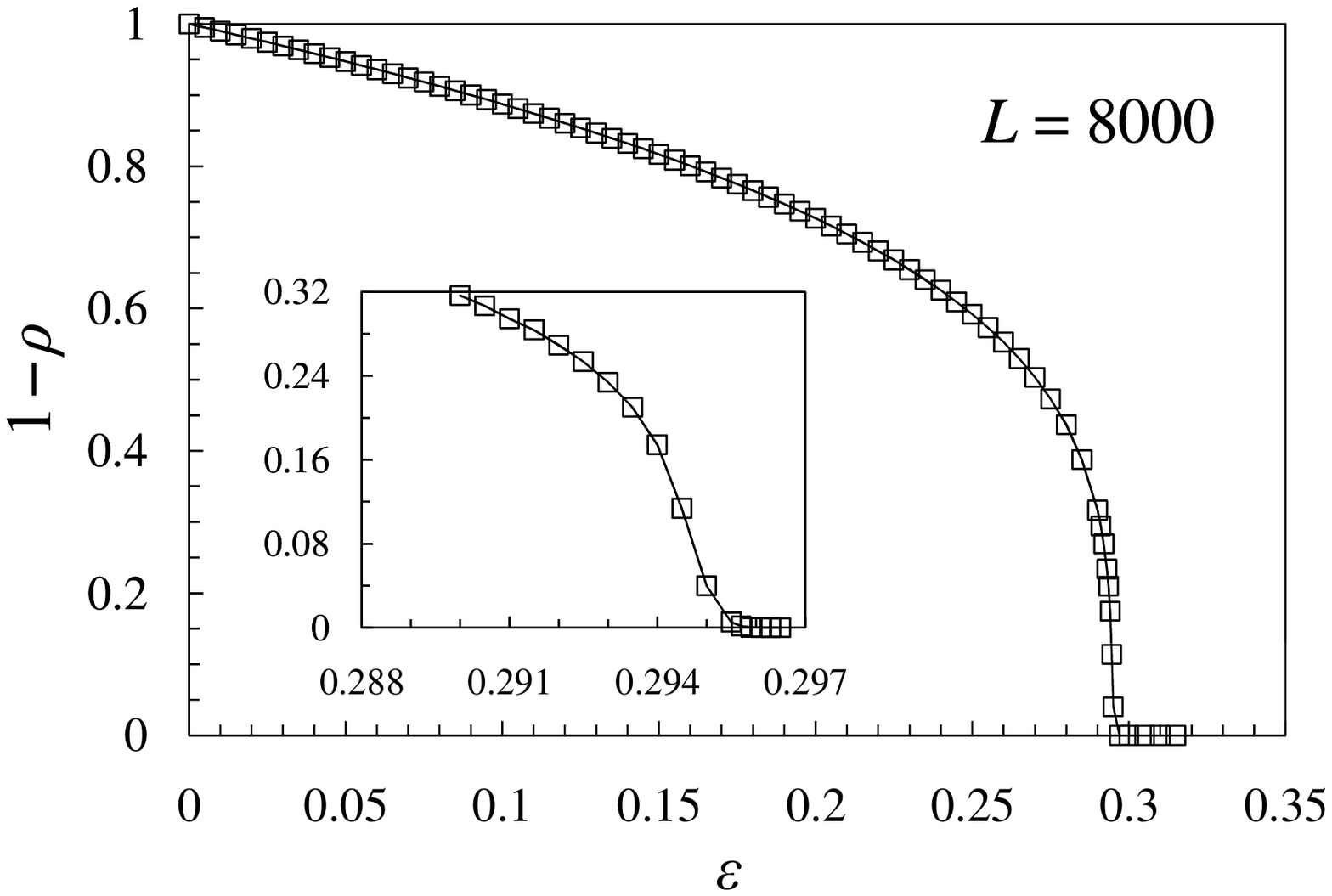}
\caption{\label{fig:rho}Stationary density of inactive cells in an automaton of $L=8000$ cells averaged over $1000$ samples. The inset shows the curve close to the critical point $\varepsilon^{*} = 0.29450(5)$ (value obtained from time-dependent simulations).}
\end{figure}

Our Monte Carlo (MC) simulations of Stavskaya's model ran as follows. For a given $\varepsilon$, the PCA is initialized with each $\eta_{i}(0)=1$, $1 \leq i \leq L$, drawn independently with probability $1/2$. Stationary state quantities, e.g. the density of active cells $\rho_{L} = L^{-1}\sum_{i}\eta_{i}$, are then sampled after the system is relaxed through $100L$ MC steps (MCS), with one MCS equivalent to a synchronous update of the states of all $L$ cells of the automaton. This amount of relaxation proved enough for our purposes. Moreover, except for the data shown in Figure~\ref{fig:rho}, our results were obtained from time-dependent simulations, so that estimates on the stationary state did not concern us much. We refer the reader to \cite{haye} for a nice exposition of the time-dependent techniques employed in what follows.

The critical behavior of the model can be determined by assuming the scaling relation
\begin{equation}
\label{scaling}
1-\rho_{L}(t;\Delta) \sim t^{-\beta/\nu_{\|}}\,\Phi(\Delta\,t^{1/\nu_{\|}},\, t^{\nu_{\perp}/\nu_{\|}}/L)
\end{equation}
close to the critical point $\varepsilon_{L}^{*}$, with $\Delta = \varepsilon -\varepsilon_{L}^{*} \geq 0$. We do not put a subscript `$L$' on $\Delta$ or the critical exponents to lighten the notation. For a very large system, relation (\ref{scaling}) becomes $1-\rho_{L}(t;\Delta) \sim t^{-\beta/\nu_{\|}}\,\Phi(\Delta t^{1/\nu_{\|}})$, with $\Phi(x \ll 1) \sim$ constant and $\Phi(x \gg 1) \sim x^{\beta}$. The investigation of the time-dependent profiles $\rho_{L}(t;\Delta)$ then allow for the simultaneous determination of $\varepsilon_{L}^{*}$ and $\delta = \beta/\nu_{\|}$, and judicious perusal of (\ref{scaling}) and derived relations furnish the other exponents.

\begin{figure}
\begin{tabular}{c}
\includegraphics[viewport=66 82 531 752, scale=0.32, angle=-90]{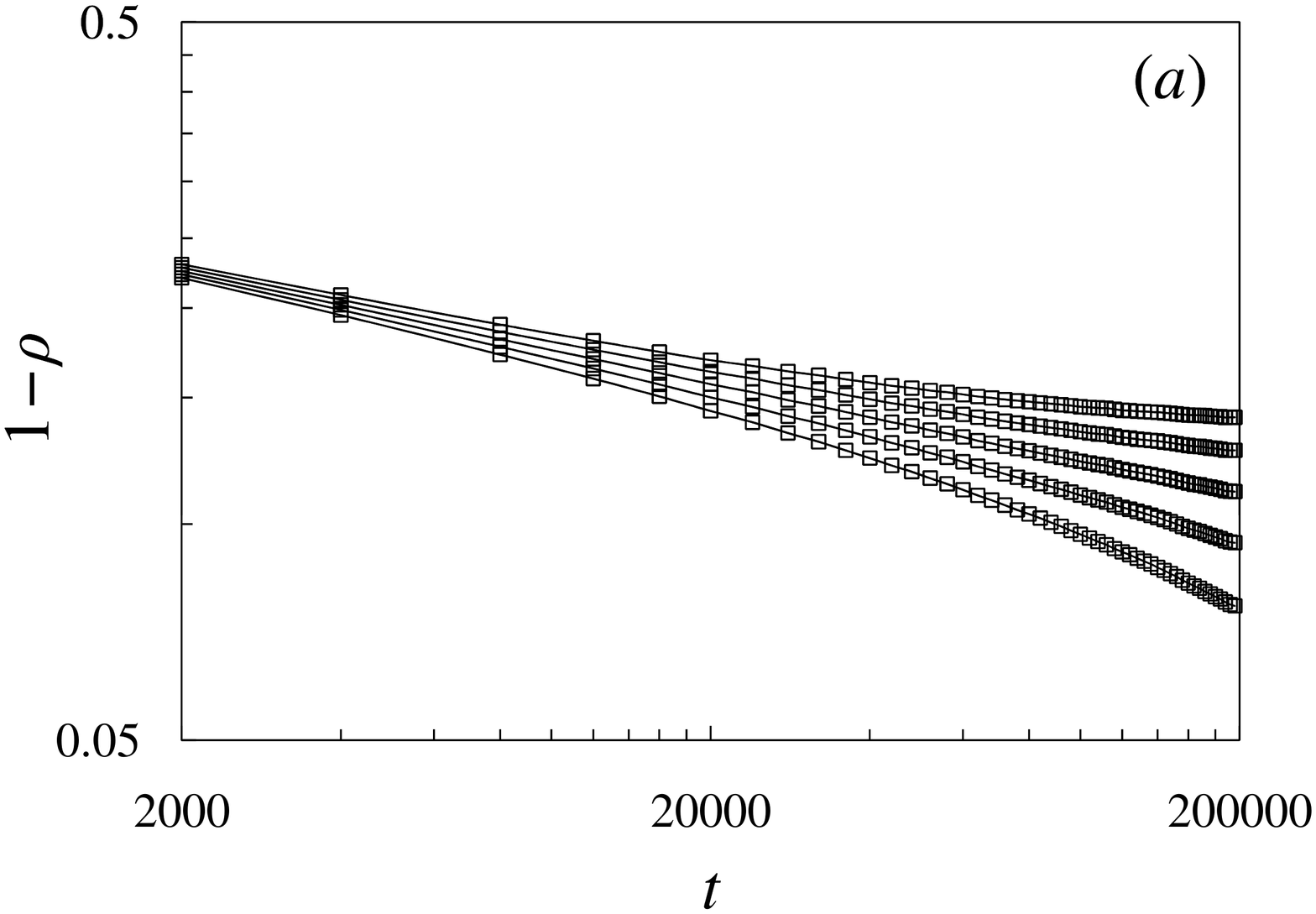} \\
\includegraphics[viewport=65 75 531 756, scale=0.32, angle=-90]{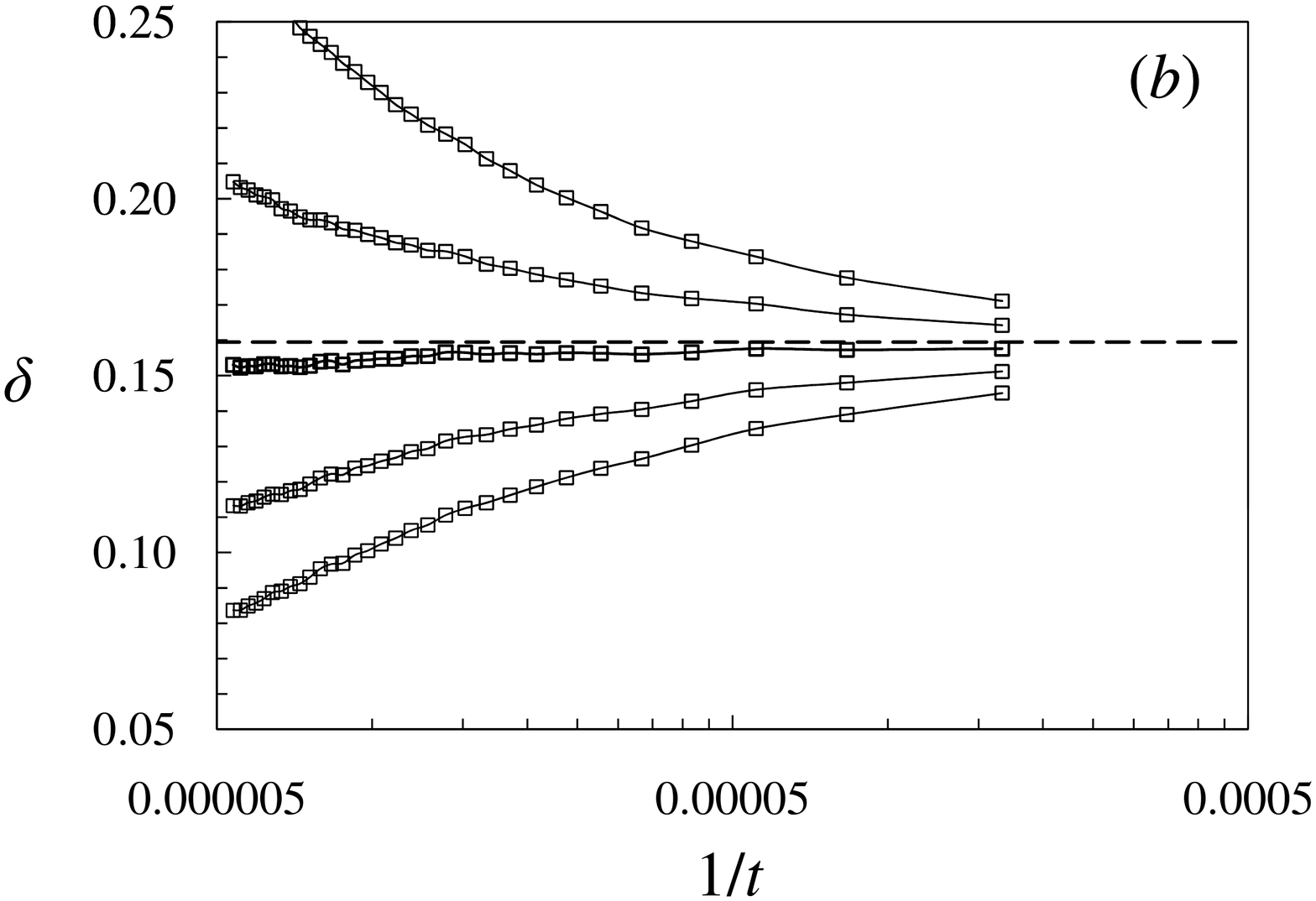}
\end{tabular}
\caption{\label{fig:delta}($a$)~Logarithmic plot of $1-\rho_{L}(t)$ with $L=20000$ cells. ($b$)~Instantaneous values of $\delta$ obtained from the curves in ($a$). In both graphs we have, from the uppermost curve downwards, $\varepsilon = 0.2943$, $0.2944$, $0.2945$, $0.2946$, and $0.2947$. From these curves we estimated $\varepsilon^{*} = 0.29450(5)$ and $\delta = 0.155(5)$. The dashed line in panel ($b$) indicates the best value available for $\delta_{\rm DP}$.}
\end{figure}

Figure~\ref{fig:rho} displays the density profile $\rho_{L}$ for an automaton of $L=8000$ cells in the stationary state. We actually plot the density of inactive cells $1-\rho_{L}$ instead, because it is this quantity that enters the scaling relation (\ref{scaling}). The steep transition about $\varepsilon_{L}^{*}$ anticipates a small value for the exponent $\beta$. To estimate $\varepsilon^{*}$ more precisely, we plot $1-\rho_{L}(t)$ close to $\varepsilon \simeq 0.294$ for some large $L$. On the critical point, $1-\rho_{L}(t) \sim t^{-\delta}$ and we can estimate $\delta$ by plotting $\log_{b}[(1-\rho_{L}(t/b))/(1-\rho_{L}(t))]$ against $1/t$ for some small $b$. Our data for $L=20000$ and $b=10$ appear in Figure~\ref{fig:delta}. From these data we could extract the estimates $\varepsilon^{*} = 0.29450(5)$ and $\delta = 0.155(5)$. Similar estimates using $4000 \leq L \leq 16000$ confirm these values. Notice that this estimate of $\varepsilon^{*}$ completely agrees with the critical point $p_{1}^{*} = p_{2}^{*} = 0.705\,489(4) = 1-\varepsilon^{*}$ found for the site DP transition in the DK PCA \cite{onody}.

\begin{figure}
\includegraphics[viewport=68 89 515 759, scale=0.32, angle=-90]{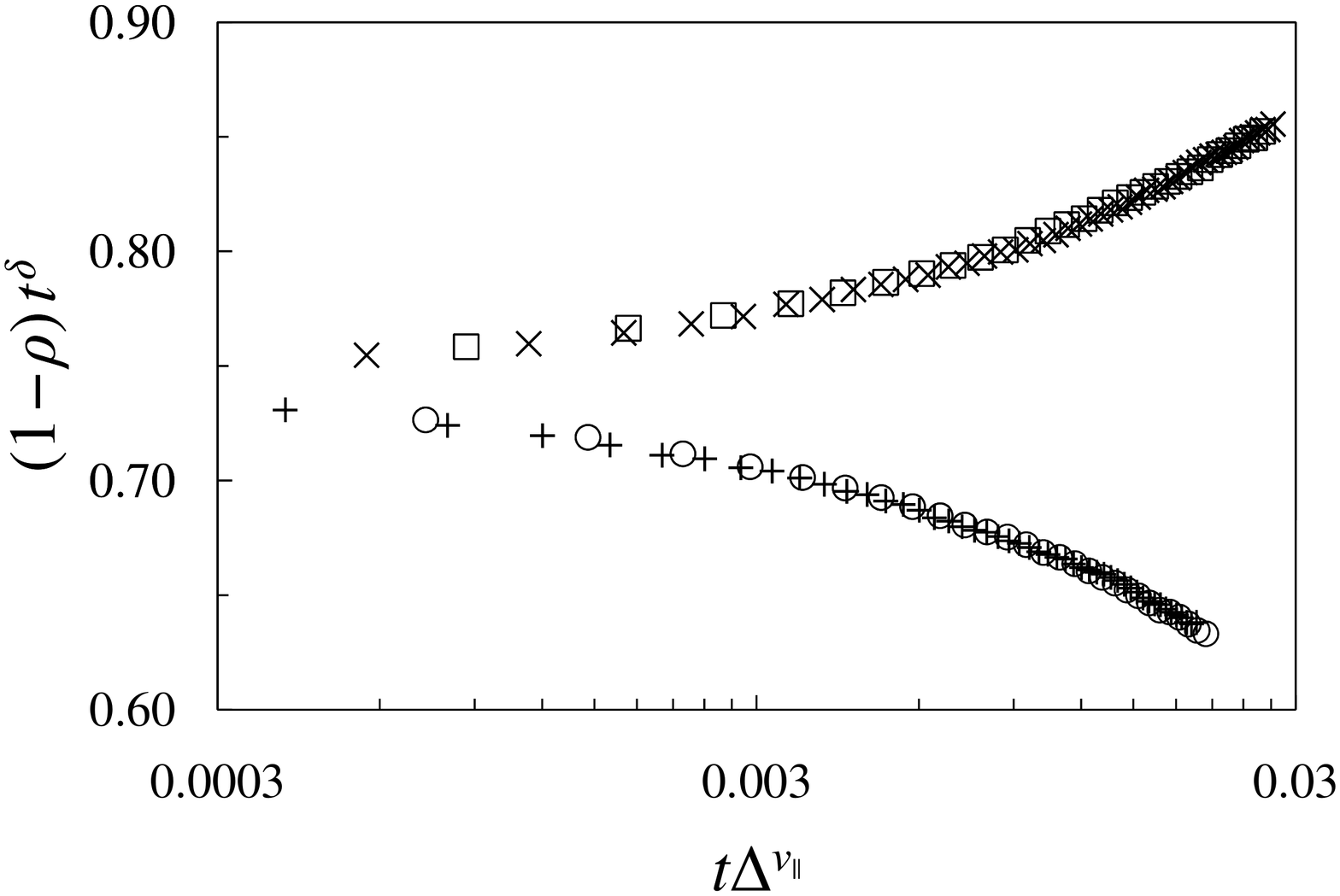}
\caption{\label{fig:nu}Data collapse of the scaled time-dependent density profiles for $\varepsilon-\varepsilon^{*} = \pm0.0001$, $\pm 0.0002$. The upper (lower) branches correspond to $\varepsilon < \varepsilon^{*}$ ($\varepsilon > \varepsilon^{*}$). The best data collapse was obtained with the central values $\varepsilon^{*}=0.29451$, $\delta=0.157$, and $\nu_{\|}=1.73$.}
\end{figure}

The exponent $\nu_{\|}$ can be obtained by plotting $t^{\delta}(1-\rho_{L}(t))$ versus $t \Delta^{\nu_{\|}}$ and tuning $\nu_{\|}$ to achieve data collapse with different $\Delta$. The collapsed curves shown in Figure~\ref{fig:nu} were obtained with a combination of central values $\varepsilon^{*}=0.29451$, $\delta=0.157$, and $\nu_{\|}=1.73$. We found it hard to discern values of $\nu_{\|}$ by less than $\pm 0.02$. Otherwise, we found the data collapse very sensitive to the choice of $\varepsilon^{*}$; in fact, it could have been used to locate $\varepsilon^{*}$ within quite tight bounds. Combining $\delta$ and $\nu_{\|}$ furnishes $\beta = \delta \nu_{\|} = 0.27(1)$ (or $\beta = 0.268(10)$).

\begin{figure}[t]
\includegraphics[viewport=68 89 515 758, scale=0.32, angle=-90]{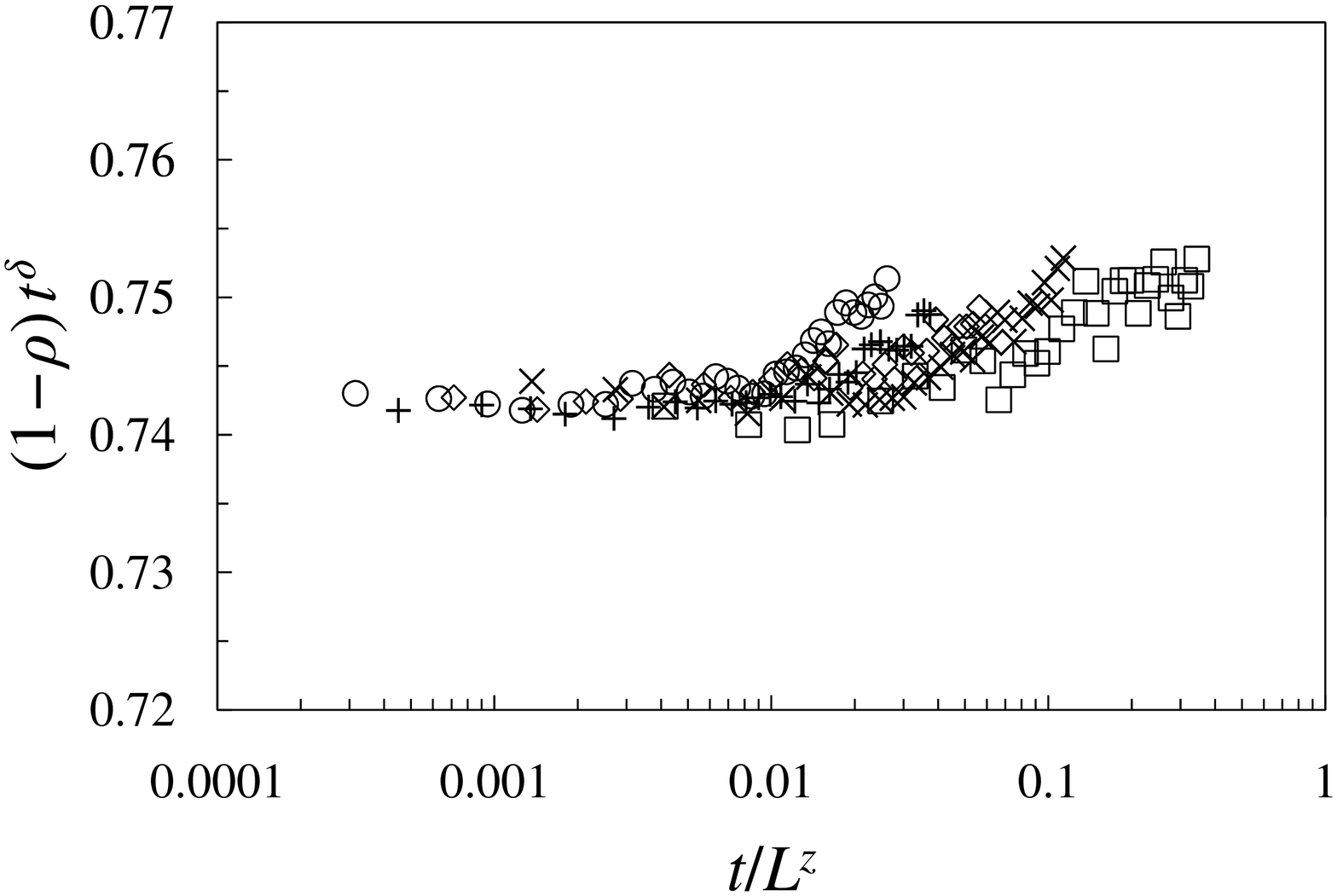}
\caption{\label{fig:ze}Finite-size data collapse of the scaled time-dependent density profiles on the critical point $\varepsilon^{*} = 0.29450$ for $4000 \leq L \leq 20000$. Best data collapse was obtained with $\delta=0.157$ (the same value as in Fig.~\ref{fig:nu}) and $z=1.6$. Notice that the data is spread over $\gtrsim 3$ decades and that the range in the $y$-axis is reasonably tight.}
\end{figure}

The third independent exponent can be obtained by plotting $t^{\delta}(1-\rho_{L}(t))$ versus $t/L^{z}$ for different $L$ and tuning $z$ until data collapse for some $z$. Since $z = \nu_{\|}/\nu_{\perp}$ by definition, this procedure also gives $\nu_{\perp}$ once $\nu_{\|}$ is known. The finite-size curves appear in Figure~\ref{fig:ze}. We found $z = 1.6(1)$. These three exponents, $\delta$, $\nu_{\|}$, and $z$, suffice to determine the universality class of critical behavior of the model, the other exponents following from well known hyperscaling relations \cite{haye}.

The best values available for $\delta$, $\nu_{\|}$, $\beta$, and $z$ for the DP process on the square lattice are $\delta_{\rm DP} = 0.159\,464(6)$, $\nu_{\|{\rm DP}} = 1.733\,847(6)$, $\beta_{\rm DP} = 0.276\,486(8)$, and $z_{\rm DP} = 1.580\,745(10)$ \cite{jensen}. Thus, within the error bars our estimates for these exponents put the critical behavior of Stavskaya's model phase transition in the DP universality class, as it would be expected on the basis of the DP conjecture.

% %% % %% % %% % %% % %% % %% % %% % %% % %% % %% % %% % %% % %% % %% % %% % %% % %% % %% % %% % 

\section{\label{summary}Summary and conclusions}

We estimated the critical point of Stavskaya's model at $\varepsilon^{*} = 0.29450(5)$ and found that the model belongs to the DP universality class of critical behavior. The value of $1-\varepsilon^{*}$ is in excellent agreement with the critical point $p_{1}^{*} = p_{2}^{*} = 0.705\,489(4)$ for the site DP transition in the DK PCA \cite{onody}.

The estimates of the critical point as well as of the critical exponents could be improved by larger simulations, but we believe that this would be superfluous, since both the location of the critical point of the model within better bounds than those provided by rigorous and mean-field analyses and the determination of its universality class of critical behavior could be established within the computational efforts reported here, namely, a few thousand hours of CPU time on Intel i7-860 processors running GCC/Linux at 2.8~GHz.

Taken together, our numerical results for Stavskaya's PCA and the rigorous results existent on its relationship with the general theory of cellular automata and percolation processes provide a reasonably complete characterization of the model. The establishment of its relationship with the DK PCA on its site DP line allows the translation of results between the two models, with potential benefits for future developments involving either model {\it per se\/} or as approximations (e.g., on coupling schemes) to models of greater complexity.

% %% % %% % %% % %% % %% % %% % %% % %% % %% % %% % %% % %% % %% % %% % %% % %% % %% % %% % %% % 

\acknowledgments

The author thanks Prof.~M\'{a}rio J. de Oliveira (IF/USP) for several helpful conversations.

% %% % %% % %% % %% % %% % %% % %% % %% % %% % %% % %% % %% % %% % %% % %% % %% % %% % %% % %% % 

\end{document}